\documentclass[review,authoryear]{elsarticle}

\usepackage[T1]{fontenc}
\usepackage[utf8]{inputenc}

\usepackage[colorlinks=true,allcolors=blue]{hyperref}
\usepackage{amsmath,amssymb,amsfonts}
\usepackage{graphicx}
\usepackage{xcolor}
\usepackage[all]{nowidow}
\usepackage[inline]{enumitem}
\usepackage[scaled]{beramono}
\usepackage{booktabs} 
\usepackage[colorinlistoftodos]{todonotes}
\usepackage[tight,footnotesize]{subfigure}
\usepackage{multirow}
\usepackage{balance}
\usepackage{listings}
\usepackage{tcolorbox}
\usepackage{cleveref}
\usepackage{soul}
\usepackage{pdflscape}
\usepackage{lineno}
\usepackage{color, colortbl}
\usepackage{geometry}
\usepackage{makecell} 

\usepackage{ltablex}
\usepackage{rotating}
\usepackage[normalem]{ulem}
\usepackage{ltablex}

\newcommand{\anhho}[1]{\textcolor{black}{{#1}}}
\newcommand{\rebuttal}[1]{\textcolor{black}{#1}}

\newcommand{\rev}[1]{\textcolor{black}{#1}}

\lstdefinelanguage{m3}{
	basicstyle=\ttfamily\scriptsize,
	keywordstyle=\bfseries,
	keywords={m3,declarations,methodInvocation},
	literate={<-}{$\leftarrow$}{1},
	tabsize=2,
	alsoletter={-}
}

\usepackage{xspace}
\newcommand*{\ie}{i.e.,\@\xspace}
\newcommand*{\eg}{e.g.,\@\xspace}

\makeatletter
\newcommand*{\etc}{%
	\@ifnextchar{.}%
	{etc}%
	{etc.\@\xspace}%
}
\makeatother
\newcommand*{\etal}{\emph{et~al.}\@\xspace}

\definecolor{darkgray}{gray}{0.78}
\definecolor{lightgray}{gray}{0.85}
\definecolor{verylightgray}{gray}{0.95}
\definecolor{mygreen}{rgb}{0,0.6,0}
\definecolor{lightgray}{gray}{0.85}

\newcommand{\ES}{\textsc{EnseSmells}\xspace}
\newcommand{\DS}{\textsc{DeepSmells}\xspace}

\newcommand{\rqfirst}{\textbf{RQ$_1$}: \emph{How do software metrics impact the performance enhancement of \ES?}}
\newcommand{\rqsecond}{\textbf{RQ$_2$}: \emph{How do various embedding techniques impact each category of code smell?}}
\newcommand{\rqthird}{\textbf{RQ$_3$}: \emph{How eﬀective is \ES compared with classical ML classifiers when utilizing only structural features?}}
\newcommand{\rqfourth}{\textbf{RQ$_4$}: \emph{How does \ES perform compared to the baselines?}}

\newtcolorbox{shadedbox}{
	drop shadow southeast,
	breakable,
	enhanced jigsaw,
	colback=white,
	boxrule=0.80pt,
	left=0.3em,
	right=0.3em,
	top=0.1em,
	bottom=0.05em
}

\modulolinenumbers[5]
\begin{document}
\begin{frontmatter}
    \title{\ES: Deep ensemble and programming language models for automated code smells detection}
\author[unimelb]{Anh Ho}
\author[hust]{Anh M. T. Bui\corref{cor1}}
\author[univaq]{Phuong T. Nguyen}
\author[gssi]{Amleto Di Salle}
    \author[unimelb]{Bach Le}
    \address[unimelb]{The University of Melbourne, Australia \\		anh.ho1@student.unimelb.edu.au, bach.le@unimelb.edu.au}
\address[hust]{Hanoi University of Science and Technology, Vietnam\\
    anhbtm@soict.hust.edu.vn}
    \address[univaq]{Universit\`a degli studi dell'Aquila, 67100 L'Aquila, Italy\\
    phuong.nguyen@univaq.it}
\address[gssi]{Gran Sasso Science Institute, Italy \\
    amleto.disalle@gssi.it}
\cortext[cor1]{Corresponding author}

\begin{abstract}
	A smell in software source code denotes an indication of suboptimal design and implementation decisions, potentially hindering the code understanding and, in turn, raising the likelihood of being prone to changes and faults. Identifying these code issues at an early stage in the software development process can mitigate these problems and enhance the overall quality of the software. Current research primarily focuses on the utilization of deep learning-based models to investigate the \anhho{contextual} information concealed within source code instructions to detect code smells, with limited attention given to the importance of structural and design-related features.
	This paper proposes a novel approach to code smell detection, \anhho{constructing} a deep learning \anhho{architecture} that places importance on the fusion of \anhho{structural} features and \anhho{statistical semantics derived from pre-trained models for programming languages}.
	We further provide a thorough analysis of how different source code embedding models affect the detection performance with respect to different code smell types. 
	\anhho{Using four widely-used code smells from well-designed datasets, our empirical study shows that incorporating design-related features significantly improves detection accuracy, outperforming state-of-the-art methods on the MLCQ dataset with with improvements ranging from 5.98\% to 28.26\%, depending on the type of code smell.}
\end{abstract}

\end{frontmatter}



\section{Introduction}\label{sec:Introduction}

\emph{Code smell}~\citep{fowler1999} is the term used to discernible traits or recurring patterns in software source code, indicating probable deficiencies or regions warranting enhancement. Code smells differ from conventional software bugs or errors in that they do not manifest as malfunctions; instead, they function as indications of latent design or implementation quandaries with the potential to precipitate issues in subsequent phases of development and maintenance. Identifying and addressing code smells is an important part of code review and maintenance, as it can help minimize technical debt and make code easier to understand, modify, and extend.

Various studies have been undertaken to investigate the smell metaphor in software engineering. These studies includes comprehensive elucidations of different kinds of code smells~\citep{zhang2011code,singh2018systematic}, methodologies for detecting such smells~\citep{ZHANG2022109737,sharma2021code} as well as quantifiable metrics that may serve as indicators for discerning bad smells~\citep{marinescu2005measurement,sharma2016designite}. 
Software metrics~\citep{lanza2006characterizing} have been widely studied to identify design defects in the source code~\citep{marinescu2004detection, munro2005product, van2007detection}.
Approaches to code smell detection, whether based on metrics or heuristics, often depended on manually setting thresholds for various measurements. 
While these methods obtain promising outcomes, crafting optimal rules or heuristics manually proves to be highly challenging~\citep{liu2015dynamic}.
Alternatively, relying solely on software metrics could result in a potential classification bias, as code snippets exhibiting distinct behaviors may share identical metrics’ value while exhibiting entirely different code smells. 

In order to bypass the need for manually designed heuristics and rules based on thresholds, machine learning techniques have been adopted to establish a sophisticated relationship between code metrics and smell labels.
A significant amount of research has focused on the application of traditional classification algorithms, such as SVM~\citep{maiga2012smurf}, Naive Bayes~\citep{khomh2011bdtex}, regression~\citep{fontana2017code}, for the detection of code smells. %
Despite their promising potential, these approaches have often overlooked the examination of code representation features.
Deep learning models have subsequently been adopted for the extraction of textual representation from source code.
Numerous studies have been dedicated to the exploration of source code mining using 
deep neural networks (DNNs)~\citep{hadj2018hybrid,liu2019deep,ho2022combining}, convolutional neural networks (CNNs)~\citep{liu2019deep,das2019detecting}, long short-term memory networks (LSTMs)~\citep{ho2023fusion}, recurrent neural networks (RNNs)~\citep{sharma2021code}, to name but few. 
Prior studies have primarily regarded source code as textual data and applied natural language processing techniques for text mining. 
However, it is evident that there exists syntactical distinctions between source code and natural language.
Approaching source code as text for mining purposes and overlooking the semantic intricacies within its underlying structures (e.g., nesting control), poses the potential of failing to preserve the intended meaning.
Moreover, code smells are frequently linked to distinct symptoms that manifest in control and data flows. 

In this paper, we propose a holistic approach  to 
code smell detection, focusing 
on assessing the effectiveness of incorporating both \anhho{statistical} semantic structures and design-related features to capture the relationship between various types of smells and source code.
In particular, we conceptualize \ES for code smell detection on top of a two-tier deep learning architecture. \anhho{The first tier focuses on acquiring \anhho{statistical} semantic representations of code by integrating \DS~\citep{ho2023fusion}, including \anhho{CNNs} and \anhho{LSTMs}, as an adapter layer to extract code-smell-specific features from code embeddings generated by pre-trained models such as Code2Vec, CodeBERT, and CuBERT. Furthermore, as code smell symptoms are often linked to specific design metrics, the second tier employs deep neural networks (DNNs) with a single adaptive layer to emphasize critical metrics. This tier is designed to learn from selected features, preserving the relationship between code smell symptoms and these metrics. The outputs from both tiers are subsequently concatenated and fed into an Imbalanced Deep Neural Networks for a more effective imbalanced classification.}

In summary, this paper extends our previous work~\citep{ho2023fusion} with the following enhancements:
\begin{itemize}
	\item \rebuttal{We enhance the approach presented in the conference version by introducing a structural module designed to automatically learn the relationships between different code smells and software metrics.}
	\item We extend the scope of experiments with a three-fold objective:
	\begin{enumerate}
		\item Investigating the impact of adjusting various code embedding techniques on prediction performance, considering four types of code smells (\ie Long Method, Feature Envy, Data Class, God Class). 
		This paper broadens the scope of code representation by exploring multiple code embedding approaches such as Code2Vec, CodeBERT, and CuBERT.
		\item Examining whether the combination of code representation vectors and hand-crafted metrics can effectively contribute to the detection of code smells.
		\item Demonstrating the effectiveness of \ES compared to state-of-the-art studies.
	\end{enumerate}
	\item We publish the packages developed alongside the metadata processed in this paper to promote future research.\footnote{\url{https://github.com/brojackvn/JSS-EnseSmells}}
\end{itemize}


Section~\ref{sec:Background} introduces the background related to the overview of code smells, the survey of software metrics for code smell detection, as well as pre-trained programming language models. It also provides an overview of related work in our research.
Section~\ref{sec:ProposedApproach} presents the proposed \ES model for code smell detection. Experimental settings, including the benchmark datasets, the employed evaluation metrics, and 
our evaluation plan, are specified in Section~\ref{sec:Evaluation}.
Section~\ref{sec:Results} discusses the results and address potential threats to the validity of our proposed approach in Section~\ref{sec:threat}.
Finally, Section~\ref{sec:Conclusion} concludes the paper with insights into future directions.


\section{\rebuttal{Background and Related Work}} \label{sec:Background}
	This section offers a brief overview of the key elements in code smell detection. Initially, we explore different code smells that have garnered attention in recent times. We also provide an overview of software metrics, considering their relevance in the study of code smells. Following this, our focus shifts to deep learning architectures, emphasizing their role in code representation and their importance in the identification of code smells.
	
	\subsection{Categorization of Code Smells}
	\label{sec:code-smells}
	\rebuttal{Fowler introduced the concept of {\it code smells} to describe potential issues within software source code that require refactoring~\citep{fowler1999}.}
	Software developers have identified a \rebuttal{wide range} 
	of code smells which can be classified into categories such as {\it implementation}, {\it design} and {\it architecture} depending on \rebuttal{their granularity level and the overall impact on the source code}~\citep{fowler1999,suryanarayana2014refactoring}.
	Implementation smells emerge at a detailed level including {\it long method, long parameter list, complex conditional}, among others~\citep{fowler1999}. 
	Design smells \rebuttal{refer to violations} of object-oriented design principles, such as abstraction and coupling, and are exemplified by issues like {\it god class, data class}, and {\it multifaceted abstraction}, among others.~\citep{suryanarayana2014refactoring}.
	Architectural smells, including instances like {\it god component and scattered functionality}, manifest at a \rebuttal{high level} of granularity, spanning across multiple components and affecting the entire system~\citep{garcia2009identifying}.
	
	In this study, we focus on four types of code smells including {\it feature envy, long method, data class} and {\it god class}
	as they have been widely used~\citep{azeem2019machine,sharma2018survey}. 
	Moreover, we utilized the MLCQ dataset containing these four smells as the benchmark~\citep{madeyski2020mlcq}.
	\rebuttal{A detailed explanation of these code smells is provided below.}
	
	\textbf{Feature Envy.} 
	Beck and Fowler introduced the concept {\it feature envy} to describe a scenario where a method accesses the data of another object more than its own data~\citep{fowler1999}.
	\rebuttal{This kind of code smell indicates misplaced methods,  typically characterized by frequent interactions with the attributes and methods of other classes.}
	
	\textbf{Long Method.} 
	A method that takes on too many responsibilities is regarded as a sign of the {\it long method}.
	This 
	can be characterized by source code complexity measurements such as lines of code (LOCs), Halstead's metrics~\citep{curtis1979measuring}, McCabe Cyclomatic~\citep{mccabe1976complexity}.
	
	\textbf{God Class.} 
	A {\it god class}, also referred to as a {\it Blob}, is characterized as a class that dominates a system's functionality by taking on too many responsibilities.
	This is reflected in its extensive number of attributes, methods and interactions with data classes.
	Similar to {\it long method}, this code smell is also distinguished by a high number of lines of code and a complex vocabulary.
	
	\textbf{Data Class.} %
	The code smell {\it data class} describes a class in a software system that mainly functions as a repository for data, offering limited functionality.
	These classes serve primarily as data holders, while the logic related to the data is often scattered across the code-base. This dispersion can result in maintenance difficulties and increase the error-proneness~\citep{fowler1999}.

	\color{black}

	\subsection{\rebuttal{Software Metric-based Approaches for Code Smell Detection}}
	
	\label{sec:CodeMetrics}

	\rebuttal{A significant amount of research has utilized software metrics to detect code smells.}
	These studies are commonly known as {\it metric-based approaches} and/or {\it rule/heuristic-based approaches}~\citep{marinescu2004detection,marinescu2005measurement,macia2010defining,vidal2015jspirit,lanza2006characterizing}.
	Marinescu et al. introduced a detection strategy based on design metrics including {\it Weighted Method Count} (WMC), {\it Tight Class Cohesion} (TCC) and {\it Access To Foreign Data} (ATFD) to identify the {\it god class} code smell~\citep{marinescu2004detection}.
	This approach was further expanded to incorporate various formulas based on source code complexity and design metrics, enabling the detection of ten different code smells~\citep{marinescu2005measurement}. 
	Lanza and Marinescu combined metric-based formulas with thresholds to detect 11 anti-patterns~\citep{lanza2006characterizing}.
	They proposed heuristics by logically combined metric-threshold pairs using logical operators to define detection rules.
	These heuristics have been adopted inside the {\it InCode}~\citep{marinescu2010incode} package as an Eclipse-plugin.
	Sales et al. have proposed a dependency based approach, called \textit{JMove} to identify {\it Feature Envy}~\citep{sales2013recommending}. 
	This approach involves defining metrics to quantify the similarity between the dependencies created by the considering method and those of all methods in dependent classes.
	\rebuttal{Likewise, numerous tools and methods have been proposed to detect different kinds of code smells including CCFinder~\citep{kamiya2002ccfinder}, SpIRIT~\citep{vidal2015jspirit}, DECOR~\citep{moha2009decor}, among others.} While these studies have yielded encouraging outcomes in detecting common code smells in source code, further extensive research is required to reach a level of maturity. 
	
	\rebuttal{
		An important challenge in rule- or heuristic-based approaches is the determination of appropriate metric thresholds. Defining these thresholds manually within smell detection algorithms represents a considerable difficulty for software engineers~\citep{liu2015dynamic}.
		Machine learning-based methods have drawn considerable interest from researchers as a solution to this challenge~\citep{maiga2012support, arcelli2016comparing,azadi2018poster}. 
		These methods involve training algorithms to capture and learn the complex relationships between source code features and code smell categories.
		Maiga et al. proposed SVMDetect based on the Support Vector Machine for code smell detection~\citep{maiga2012support}. They employed an input vector encompassing 60 structural metrics for each class to detect four well known code smells including God Class, Functional Decomposition, Spagheti Code and Swiss Army Knife. 
		Fontana et al. investigated 16 different machine learning algorithms along with their boosting variants to detect four specific code smells: data class, god class, feature envy and long method~\citep{arcelli2016comparing}. 
		The experiments were carried out using independent metrics at various levels of granularity including method, class, package and project, showcasing the potential of machine learning-based approach comparing to rule-based ones. 
		However, the ambiguity in detecting code smells from a given code snippet arises because various code smells may be associated with the same metrics~\citep{sharma2021code}. This necessitates distinct characteristics to effectively differentiate one type of code smell from others.
	}

	\subsection{\rebuttal{Deep Learning-based Approaches for Code Smell Detection}
	}
	\label{sec:EmbeddingTechniques}
	\rev{A growing body of research has shifted towards applying deep learning models to capture deeper features from source code and to learn the complex relationships between code snippets and code smell classes~\citep{liu2019deep,sharma2021code,ho2023fusion}. 
		Building on this work, we introduced \DS as an effective tool for code smell detection, leveraging diverse deep learning methods to identify patterns and semantic features in code snippets, outperforming previous approaches~\citep{ho2023fusion}. Another research direction has focused on utilizing code embeddings, marking significant advancements in applying NLP techniques to software engineering~\citep{von2022validity}. 
		Recently, there has been a shift towards Transformer-based models (e.g., BERT~\citep{devlin2018bert}, RoBERTa~\citep{liu2019roberta}, CodeBERT~\citep{feng2020codebert}, among others). These models leverage large datasets and Transformer architectures with attention mechanisms to capture various distinctive features of code, a process commonly referred to as capturing semantics. Prior studies on code smell detection have successfully applied these models to automate tasks requiring an understanding of program semantics~\citep{nguyen2024encoding}.
		Kova{\v{c}}evi{'c} et al. explored the effectiveness of code embedding models, such as \textit{code2vec}, \textit{code2seq}, and CuBERT, in comparison to traditional code metrics for detecting God Class and Long Method code smells~\citep{kovavcevic2022automatic}. 
	} 

		\rev{
			Motivated by these advancements, we aim to leverage pre-trained code and language models for code smell detection, focusing on extracting additional features. We refer to these as \textit{statistical semantic} features to distinguish them from literal semantics (\eg operational semantics in programming languages) and to highlight their derivation from statistical patterns learned by these models. The concept of ``statistical semantics'' was previously introduced in the book Statistical Semantics~\citep{sikstrom2020statistical}, where it refers to understanding and modeling the meaning of words and concepts based on their statistical co-occurrence in large text corpora. This approach assumes that meaning emerges from distributional patterns in natural language rather than from predefined dictionaries or human-labeled definitions. Extending this concept to the domain of source code analysis, we apply the term ``statistical semantics'' to describe the extraction of meaning from source code using statistical patterns. Pre-trained models such as code2vec, CodeBERT, and CuBERT are trained on a large corpora of source code, leveraging these patterns to derive semantic representations of code snippets. Similar to natural language models, these code models capture both syntactic and semantic patterns, though their effectiveness varies. Recent research~\citep{ma2024unveiling} reveals that different models exhibit distinct capabilities in encoding syntactic and semantic properties of code. Thus, we adopt ``statistical semantics'' to describe this feature, as it reflects the statistical nature of meaning extraction in source code.
		}
		
		To harness the benefits of both code metrics and statistical semantic features, we integrate these two types of features in our proposed approach, detailed in Section~\ref{sec:ProposedApproach}.

\section{The Proposed Approach}
\label{sec:ProposedApproach}

Building upon our prior research, namely \textsc{DeepSmells}~\citep{ho2023fusion}, where we concentrated on extracting unique features and patterns from source code, our current focus is on seamlessly integrating the strengths of code metrics-based and deep learning approaches. \ES combines the automated generation of both \textit{\rebuttal{statistical} semantic} and \textit{structural} features from source code. \rebuttal{The aim is to improve code smell detection accuracy and address the specific challenges encountered in previous methods.}

\begin{figure}[ht!]
	\centering
	\includegraphics[width=\columnwidth]{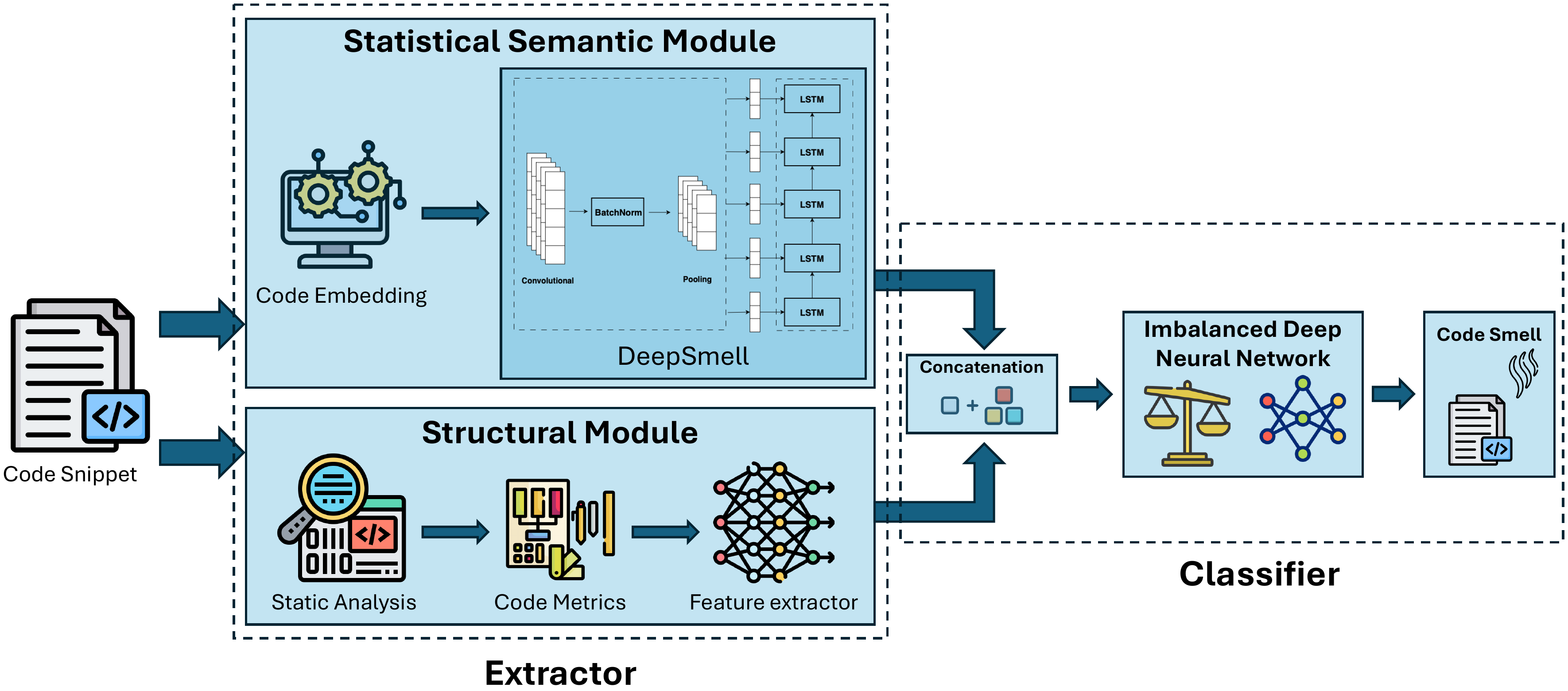}
	\caption{The overall architecture of \ES.}
	\label{fig:overallJSS}
\end{figure}

\rebuttal{As illustrated in Figure~\ref{fig:overallJSS}, our approach comprises two primary components, including \textit{Extractor} and \textit{Classifier}. The \textit{Extractor} component is designed to maximize the capture of relevant features from code snippets. Specifically, we focus on two types of features: \textit{(i)} \rebuttal{statistical} semantic features, extending our previous work~\citep{ho2023fusion}; \textit{(ii)} \rebuttal{structural features, automatically learned through a broad view of code metrics.}
	The extracted features are then concatenated and fed into the \textit{Classifier} component, where they are processed by an imbalanced DNN for classification. Once the \textit{Classifier} is constructed, with weights and biases determined, it calculates a probability for each code snippet, indicating whether it exhibits a code smell or is clean.}

\subsection{\anhho{Extractor}}
\anhho{The \textit{Extractor} includes two modules to capture distinct features of code snippets: code structure and statistical semantics.}

\subsubsection{\anhho{Statistical Semantic Module}}
\label{subsec:SemanticModule}

\anhho{This module is designed to capture statistical semantic features by leveraging code embedding models, as illustrated in Figure~\ref{fig:overallJSS}. It operates in two stages: the encoding phase and the model-building phase.
	During the encoding phase, we have experimented with three pre-trained code and language models including CodeBERT, CuBERT and {\it code2vec}. Additionally, to evaluate the effectiveness of these models compared to a natural language embedding model, we have also employed the token indexing technique to represent a source code as a vector of token indices.
}
\rebuttal
{
	Unlike simple index vectors that provide minimal contextual information about source code, pre-trained code embedding models are specifically designed to represent programming languages and are trained on extensive and complex datasets. These models extract statistical and semantic features from code snippets, enabling a more comprehensive and nuanced understanding of the underlying code semantics.
}

\begin{itemize} 
	\item \textbf{\textit{Token-Indexing}} involves mapping tokens to integers to convert token sequences into integer vectors. To do so, we conducted several steps, including:
	\textit{(i)} embedding code tokens with a tokenizer\footnote{Available at: \url{https://github.com/dspinellis/tokenizer}};
	\textit{(ii)} computing statistical information on the sample lengths, excluding those that deviate by more than one standard deviation from the mean; 
	\textit{(iii)} padding sequences with zeros to align with the longest input. This method was previously utilized in our prior work, \DS~\citep{ho2023fusion}.
	
	\item \textbf{\textit{CodeBERT}} formats input as two segments joined by special tokens: $[CLS], w_1, w_2, ..., w_n,$ $[SEP], c_1, c_2, ..., c_m, [EOS]$. The first segment generally contains natural language text, such as function comments, while the second segment consists of code tokens. The $[CLS]$ token at the beginning represents the entire sequence, aiding in classification tasks~\citep{feng2020codebert}. CodeBERT produces two types of outputs: \textit{(i)} contextual vector representations for each token in both the text and code segments, and \textit{(ii)} an aggregated vector for the entire sequence represented by the $[CLS]$ token.
	
	\item \textbf{\textit{CuBERT}} uses a custom Java tokenizer developed by Kanade \etal~\citep{kanade2020learning} to preprocess code snippets before feeding them into the CuBERT model. CuBERT processes individual lines of code as input, transforming each line into a 1024-dimensional embedding derived from the special $[CLS]$ token~\citep{devlin2018bert}. To encode each method in the input source code, we compute embeddings for each line of code within the method body and then aggregate them by summing the resulting vectors. Similarly, for class-level embeddings, we apply this process across all methods in the class and calculate the sum of their embeddings. This method of summing embeddings has been shown to deliver consistently strong performance in recent studies~\citep{kovavcevic2022automatic}.
	
	\item \textbf{\textit{code2vec}} generates fixed-length embeddings from code snippets, using method bodies as inputs and producing corresponding tags as outputs~\citep{alon2019code2vec}. To generate the code embedding, we apply {\it code2vec} to represent each method as a 384-dimensional vector through the following steps: \textit{(i)} extracting AST paths from each method using a Java path extractor, and \textit{(ii)} feeding these paths into a pre-trained {\it code2vec} model, omitting the \texttt{softmax} layer to obtain the code vector. 
	\rebuttal{To compute the embedding for a class, we treat it as a collection of methods. The representation vector of a class is computed as the average of all its method embeddings. This approach was also be employed in previous studies~\citep{compton2020embedding}.}
	
\end{itemize}

\rebuttal{To address code smell detection, we utilize our previous model, \textsc{DeepSmells}, which has demonstrated effectiveness in this task. In the model-building phase, the \anhho{code embedding} is processed through \textsc{DeepSmells}. 
	The model initially processes the embedding through convolutional blocks, which automatically extract distinctive features. While stacking multiple convolutional blocks can capture increasingly complex and abstract features, including high-level token relationships, deeper networks often encounter a degradation problem where accuracy plateaus and then declines as the network depth grows~\citep{he2016deep}. To mitigate this issue, we use two convolutional blocks, as outlined below.}

\begin{itemize}[leftmargin=*]
	\item The first convolution block consists of a 1D-CNN (\texttt{torch.nn.Conv1d}) with 16 filters with a kernel size $k$ which specifies the length of the 1D convolution window. 
	We experiment different values of $k$ (i.e., $k = 3,4,5,6,7$) to evaluate the effectiveness of the convolution layer.
	This layer employs a \texttt{ReLU} activation function, and the remaining parameters are kept as defaults, following by a 1D-Batch Normalization (\texttt{torch.nn.BatchNorm1d}) to accelerate the network training and to reduce internal covariate shift~\citep{ioffe2015batch}. The obtained feature map is then passed to a \texttt{MaxPooling} layer (\texttt{torch.nn.MaxPool1d}) by a factor of size three to reduce the spatial dimension.
	\item The second convolution block is similar to the first one, excepting that we use 32 filters. 
\end{itemize}

Once the output from the convolutional neural networks has been obtained, it is fed into an LSTM network to preserve the meaning and context of the data. The \texttt{torch.nn.LSTM} function is used to configure the LSTM network, with the input size of each LSTM unit being based on the initial size of the initial embedding source code vector. The number of LSTM units in the network is also set to 32 to match the 32 filters in the second convolution block.
Additionally, we adopt a Bi-LSTM architecture to capture long-term dependencies in sequential data, processing it in both forward and backward directions. This architecture consists of two separate LSTM layers: one processes the input sequence in the forward direction, while the other processes it in reverse. The hidden state at each time step is the concatenation of the forward and backward hidden states, enabling the model to incorporate information from both past and future contexts. We maintain the same configuration as before but set the \texttt{bidirectional} hyperparameter to \texttt{True}. 
The final hidden state from this network serves as the \textit{\anhho{statistical} semantic features} of this module.

\subsubsection{Structural Module}
\rebuttal{
	Previous studies have demonstrated the effectiveness of source code metrics in detecting code smells using rule- or heuristic-based approaches~\citep{marinescu2010incode,sales2013recommending,vidal2015jspirit}. In this research, we also take into consideration source code metrics as structural features. However, instead of relying on threshold-based rules, we propose a deep learning model to capture non-linear relationships between code metric vectors and different categories of code smells (Figure~\ref{fig:overallJSS}).
}

\begin{table}[h!]
	\caption{The class-level source code metrics.}
	\label{table:ClassLevelMetric}
	\scriptsize
	\centering
	\vspace{0.1cm}
	\begin{tabular}{llll}
		\Xhline{1pt}
		AnonymousClassesQty    & AssignmentsQty        & TotalFieldsQty       & LogStatementsQty \\
		InnerClassesQty        & ReturnQty             & CisibleFieldsQty     & Modifiers              \\
		AbstractMethodsQty     & LoopQty               & VariablesQty         & TCC              \\
		PrivateMethodsQty      & UniqueWordsQty        & TryCatchQty          & LCC              \\
		ProtectedMethodsQty    & PrivateFieldsQty      & LambdasQty           & WMC              \\
		PublicMethodsQty       & ProtectedFieldsQty    & ParenthesizedExpsQty & LOC             \\
		DefaultMethodsQty      & PublicFieldsQty       & NumbersQty           & LCOM              \\
		StaticMethodsQty       & DefaultFieldsQty      & ComparisonsQty       & RFC              \\
		FinalMethodsQty        & StaticFieldsQty       & MaxNestedBlocksQty   & CBO              \\
		SynchronizedMethodsQty & FinalFieldsQty        & MathOperationsQty    & DIT        \\
		TotalMethodsQty        & SynchronizedFieldsQty & StringLiteralsQty    & NOSI             \\
		\Xhline{1pt}
	\end{tabular}
\end{table}

\begin{table}[ht!]
	\caption{The method-level source code metrics.}
	\label{table:MethodLevelMetric}
	\scriptsize
	\centering
	\vspace{0.3cm}
	\begin{tabular}{llll}
		\Xhline{1pt}
		LOC         & VariablesQty                   & TryCatchQty          & AnonymousClassesQty \\
		RFC         & ParametersQty                  & ParenthesizedExpsQty & InnerClassesQty     \\
		CBO         & MethodsInvokedQty              & StringLiteralsQty    & LambdasQty          \\
		WMC         & MethodsInvokedLocalQty         & NumbersQty           & UniqueWordsQty      \\
		LogStatementsQty  & MethodsInvokedIndirectLocalQty & AssignmentsQty       & Modifiers           \\
		Constructor & LoopQty                        & MathOperationsQty    &     \\
		ReturnsQty  & ComparisonsQty                 & MaxNestedBlocksQty   &            \\
		\Xhline{1pt}
	\end{tabular}
\end{table}

\rebuttal{
	Code snippets are first analyzed to compute software metrics. Arcelli et al. synthesized and classified these metrics into different aspects, including size, complexity, cohesion, coupling, encapsulation, and inheritance~\citep{arcelli2016comparing}. Tables~\ref{table:ClassLevelMetric} and~\ref{table:MethodLevelMetric} provide a summary of commonly used metrics in prior research, including 44 class-level metrics and 26 method-level metrics\footnote{Only the metric abbreviations are shown here; the full names and detailed descriptions are provided in the Appendix.}. 
	These metrics cover multiple dimensions of code structure, complexity, and object-oriented design, which have been demonstrated in numerous studies to have a strong correlation with different types of code smells~\citep{marinescu2005measurement,lanza2006characterizing,marinescu2010incode,maiga2012support}.
}
\rebuttal{
We utilized the \texttt{CK} tool\footnote{Available at: \url{https://github.com/mauricioaniche/ck}} to calculate the code metrics for each code snippet, based on whether it is at the class or method level.
Before feeding the code metric vector into a deep learning model, some pre-processing steps have been conducted as follows.
\begin{itemize}
\item Categorical value metrics are encoded using label encoding.
\item Metrics with constant values are removed as they are ineffective for making distinctions.
\item In cases where a metric value could not be calculated for a specific code sample, we used k-Nearest Neighbors from the \texttt{scikit-learn} library\footnote{Available at: \url{https://scikit-learn.org/}} to impute the missing value. However, if the number of missing data samples for a given metric exceeds 5\% of the total samples, we exclude this metric~\citep{schafer1997analysis,bennett2001can}.
\item Finally, all metric features are normalized using \texttt{StandardScaler} from the \texttt{scikit-learn} library.
\end{itemize}
These pre-processing steps ensure the data is properly formatted for the final stage, where we apply a deep neural network with a single adaptive layer. This layer is both efficient and convenient, as it dynamically adjusts the number of hidden nodes, unlike traditional neural networks, where parameters are tuned through iterative processes. The adaptive layer assigns higher weights to the importance of each metric and maps them into a lower-dimensional latent space~\citep{xu2019software}. This approach effectively captures the key \textit{structural features} necessary for code smell detection.
}

\vspace{-0.1cm}
\subsection{Classifier}
\anhho{The \textit{Classifier} component combines both structural and statistical semantic features from the source code using an ensemble approach (Figure~\ref{fig:overallJSS}).} 

\rebuttal{
In our previous work~\citep{ho2023fusion}, we relied solely on the token-indexing technique to encode code snippets, paired with a simple deep learning model for classifying smelly and non-smelly source code. In this study, we aim to employ a more complex model architecture that learns a richer representation of code snippets, incorporating both statistical semantic and structural features for more effective code smell detection.
Two embedding vectors resulting from the {\it Extractor} component are concatenated using the \texttt{torch.cat} operator of \texttt{PyTorch}. The ensemble vector is then fed as input into the Deep Imbalanced Neural network for classification.
}
\rebuttal{This network comprises two hidden layers using \texttt{ReLU}~\citep{nair2010rectified} as the activation function. The output layer, designed for the binary classification task, consists of a single node with a \texttt{Sigmoid} activation function.  The optimal number of hidden nodes is determined empirically, allowing the model to adapt to different datasets.
Due to a significant class imbalance, with the number of smelly code snippets being much lower than non-smelly ones, the model may develop a bias toward the majority class, despite the minority class being more important. To address this issue, we introduce a sensitivity weight into the binary cross-entropy loss function.
\begin{equation}
\label{eq:loss-function}
\mathcal{L}(x_i) = \beta \hat{y_i} log(y_i) + (1-\hat{y_i}) log(1-y_i)
\end{equation}
where $\beta$ is the sensitive weight, $\hat{y_i}$ is the actual label of input $x_i$ and  $y_i$ is the model's prediction for input $x_i$.}
This weight adjustment enables us to fine-tune the importance of the minority class in the classification process. To determine the optimal weight, we conduct a series of experiments comparing the performance of the binary cross-entropy with $\beta$ setting to 1 against weighted binary cross-entropy with varying weight values for $\beta$, such as 2, 4, 8, 12, 32, and 84. The results of these comparisons help us select the best hyper-parameter for our proposed method.

During the training and testing phase, we trained our model using a mini-batch size of 128 and set the learning rate to 0.025. Training was performed over 85 epochs using SGD to optimize the weighted loss function.


\section{\rebuttal{Experimental Settings}}
\label{sec:Evaluation}
We ran the evaluation with Google Colab Pro\footnote{Available at: \url{https://colab.research.google.com/}} using a virtual machine equipped with an Intel Xeon CPU with 2 vCPUs (virtual CPUs) and 13GB of RAM. 
The virtual machine was equipped with a NVIDIA Tesla T4 GPU with 16GB of VRAM to accelerate the deep learning computations. To evaluate 
\ES, we re-ran all the baselines using the same benchmark datasets introduced in Section~\ref{sec:BenchmarkDataset}. To ensure a robust evaluation, we initially split the dataset into 80\%:20\% for training and testing, respectively. Moreover, to mitigate any potential bias and variance related to the test set resulting from the dataset split, stratified 5-fold cross-validation was applied, resembling traditional k-fold cross-validation but preserving the class distribution within each split. 

\subsection{Research questions} \label{sec:ResearchQuestions}

\rqfirst~\anhho{The objective of this research question is to investigate the impact of adding structural modules to \DS, forming \ES. To achieve this, we compare \ES and \DS under the same configurations for the statistical semantic module.}



\rqsecond~\anhho{The objective of this research question is to investigate code embedding techniques, each characterized by unique architectures that capture distinctive features associated with individual smells. To achieve this, we compare the performance of the aforementioned code embeddings in each architecture (\ES and \DS) under each category of code smell.}


\rqthird~\anhho{The objective of this research question is to evaluate the effectiveness of \ES, which combines various features within a complex architecture, compared to traditional machine learning classifiers that rely solely on software metrics proven effective for performing on tabular data (i.e., our structural features). To achieve this, we perform a grid search on classic classifiers with configurations outlined in Table 4 to determine the optimal choice and compare \ES under its optimal configuration.}

\rqfourth~\anhho{The objective of this research question is to demonstrate the performance of \ES compared to baseline models. To achieve this, we conduct experiments with existing studies, including our work~\citep{ho2023fusion}, the state-of-the-art, ML\_CuBERT model~\citep{kovavcevic2022automatic} and 
	three baseline models~\citep{sharma2021code}. Three baselines are variants of an auto-encoder model designed to compress source code and learn salient information reflected in the reconstructed output, which are: 
	\textit{(i) AE-Dense:} An auto-encoder model that employs dense layers for both the encoder and decoder. \textit{(ii) AE-CNN:} An auto-encoder model that employs two CNN networks, one for the encoder and another for the decoder. \textit{(iii) AE-LSTM:} Similar to AE-CNN, but with CNNs replaced by LSTM networks. In all cases, the encoder and decoder layers are followed by a fully connected dense layer for classification.}

\subsection{Benchmark dataset}
\label{sec:BenchmarkDataset}

The MLCQ dataset, originally made available by an existing study~\citep{madeyski2020mlcq}, was used in our experiments. The dataset comprises 14,739 reviews related to approximately 5,000 Java code snippets gathered from 26 software developers. The reviews are categorized into four types of code smells, \ie God Class and Data Class at the class level, and Feature Envy and Long Method at the method level. 

\anhho{The initial dataset labeling follows a structured approach: each reviewer assigns exactly one severity level for each code smell in a given code snippet, choosing from four levels: \textit{none, minor, major}, and \textit{critical}. We defined the task as a binary classification for each code snippet, labeling it as either \textit{smelly} or \textit{non-smelly}. However, since each code snippet has multiple reviews, we determine the final label using \textit{majority vote} to reconcile the varying labels. Code samples are labeled as \textit{smelly} if they receive more smelly reviews (i.e., minor, major, or critical severity) than non-smelly reviews (i.e., none severity). Conversely, they are classified as \textit{non-smelly} if they have a greater number of non-smelly reviews. To maintain a clear classification, code snippets with an equal count of smelly and non-smelly reviews are subsequently removed from the dataset, as they do not provide a definitive classification and could introduce ambiguity into the binary labeling process.} 

\anhho{The resulting dataset is summarized in Table~\ref{table:BenchmarkDataset}. As shown, this classification problem exhibits a significant class imbalance, with the positive class (smelly instances) representing the minority class. The average imbalance rate across all datasets is 10.46\%.}

\begin{table}[ht!]
	\caption{Statistics of the datasets.}
	\label{table:BenchmarkDataset}
	\centering
	\scriptsize
	\vspace{0.3cm}
	\begin{tabular}{cccc}
		\Xhline{1pt}
		\textbf{Smell} & \textbf{Smell Alias} & \anhho{\textbf{\# Negative }} & \anhho{\textbf{\# Positive}}    \\ 
		\Xhline{1pt}
		Long Method    & LM                   & 1,993                 & 243                  \\
		Feature Envy   & FE                   & 2,126                 & 64                   \\
		Data Class     & DC                   & 1,838                 & 282                  \\
		God Class      & GC                   & 1,857                 & 228                  \\ 
		\Xhline{1pt}
	\end{tabular}
\end{table}

\subsection{Evaluation metrics}
\anhho{Performance metrics commonly used in classification problems include Precision (P), Recall (R), F1-Score (F1), and Accuracy. However, these metrics have specific limitations. For instance, Accuracy often performs poorly on imbalanced datasets as it disproportionately favors the majority class. Additionally, Precision, Recall, and F1-Score focus on three quadrants of the confusion matrix (true positives, false positives, and false negatives) while excluding true negatives, which can lead to incomplete interpretations. To enable meaningful comparisons, we report these metrics while also including Matthews Correlation Coefficient (MCC), which provides a more balanced evaluation by considering all quadrants of the confusion matrix. Consistent with prior work~\citep{sharma2021code, madeyski2023detecting, kovavcevic2022automatic}, these metrics are used to assess model performance.}

\textbf{Precision, Recall, and F1-Score.}
Confusion matrix is an effective means 
to evaluate any classifier with four possible outcomes including {\it true positive} (TP), {\it true negative} (TN), {\it false positive} (FP) and {\it false negative} (FN). {\it Precision} measures how many of the positive predictions made by the model are actually correct: $P = \frac{TP}{TP+FP}$; {\it Recall} counts the number of positive cases in the dataset that model can identify: $R = \frac{TP}{TP+FN}$; {\it F1-Score} represents the harmonic mean of {\it precision} and {\it recall} of the prediction model: $F1 = \frac{2*P*R}{P + R}$.

\textbf{MCC.} The metric is used 
with imbalanced classification, measuring the correlation between the predicted and actual class, which lies 
in the range $[-1,1]$, where $1$ represents a perfect prediction, and $-1$ shows a perfect negative correlation; 
An MCC equal to $0$ corresponds to a random prediction.
\begin{equation}
	MCC = \frac{TP*TN-FP*FN}{\sqrt{(TP+FP)(TP+FN)(TN+FP)(TN+FN)}}
\end{equation}

\anhho{Notably, MCC and F1-Score both reflect overall performance~\citep{sharma2021code, madeyski2023detecting}: MCC offers a balanced assessment by considering all quadrants of the confusion matrix, while F1-Score is widely used to capture the trade-off between precision and recall, offering insights into the classifier’s ability to identify positive cases.}


	\begin{table}[ht!]
		\caption{The parameter settings of the used machine learning classifiers.}
		\label{table:MachineLearningHyperparameter}
		\centering 
		\scriptsize
		\vspace{0.3cm}
		\begin{tabular}{ll}
			\Xhline{1pt}
			\textbf{Classifier} & \textbf{Hyperparameter settings} (\textit{scikit-learn} library)\\ 
			\Xhline{1pt}
			
			Naive Bayes (NB)                    & \texttt{sklearn.naive\_bayes.GaussianNB}: \\ &$var\_smoothing=[1e^{-9}, 1e^{-8}, 1e^{-7}, 1e^{-6}, 1e^{-5}]$ \\ \hline
			
			Nearest Neighbor (NN)                    & \texttt{sklearn.neighbors.KNeighborsClassifier}:\\ & $n\_neighbors=[1, 3, 5, 7, 9]$ \\ \hline
			
			Random Forest (RF)                    & \texttt{sklearn.ensemble.RandomForestClassifier}: \\&$n\_estimators=[10, 50, 100, 200]$\\& $max\_depth=[None, 10, 20, 30]$ \\ \hline
			
			Logistic Regression (LR)                    & \texttt{sklearn.linear\_model.LogisticRegression}: \\& $C\_values=[0.001, 0.01, 0.1, 1, 10, 100]$ \\ \hline
			
			Classification and                  & \texttt{sklearn.tree.DecisionTreeClassifier}: \\Regression Tree (CART)&$max\_depth\_values=[None, 10, 20, 30]$ \\&$min\_samples\_split\_values=[2,5,10]$ \\&$min\_samples\_leaf\_values=[1,2,4]$ \\&$max\_features\_values=[auto, sqrt, log2]$ \\ \hline
			
			Support Vector                    & \texttt{sklearn.svm.SVC}: \\Machine (SVM)&$kernel\_function=Gaussian RBF$ \\& $gamma\_values=[0.001, 0.01, 0.1, 1, 10, 100]$ \\& $C\_values=[0.001, 0.01, 0.1, 1, 10, 100]$        \\ \hline
			
			Back Propagation                    & \texttt{sklearn.neural\_network.MLPClassifier}: \\Neural Networks (BP)& $hidden\_layer\_size=[16, 32, 64, (32, 16)]$ \\&$learning\_rate\_init=[0.001, 0.01, 0.1]$ \\& $max\_iter=[100,200,300]$ \\& $tol\_err=[1e^{-4}, 1e^{-3}, 1e^{-2}]$ \\ 
			\Xhline{1pt}
			
		\end{tabular}
	\end{table}

\section{\rev{Experimental Results and Discussion}}\label{sec:Results}


\rebuttal{This section reports and analyzes the experimental results by answering the research questions introduced in Section~\ref{sec:ResearchQuestions}}.

\subsection{\rqfirst}

\begin{figure}[t!]
	\centering
	\includegraphics[width=0.80\linewidth]{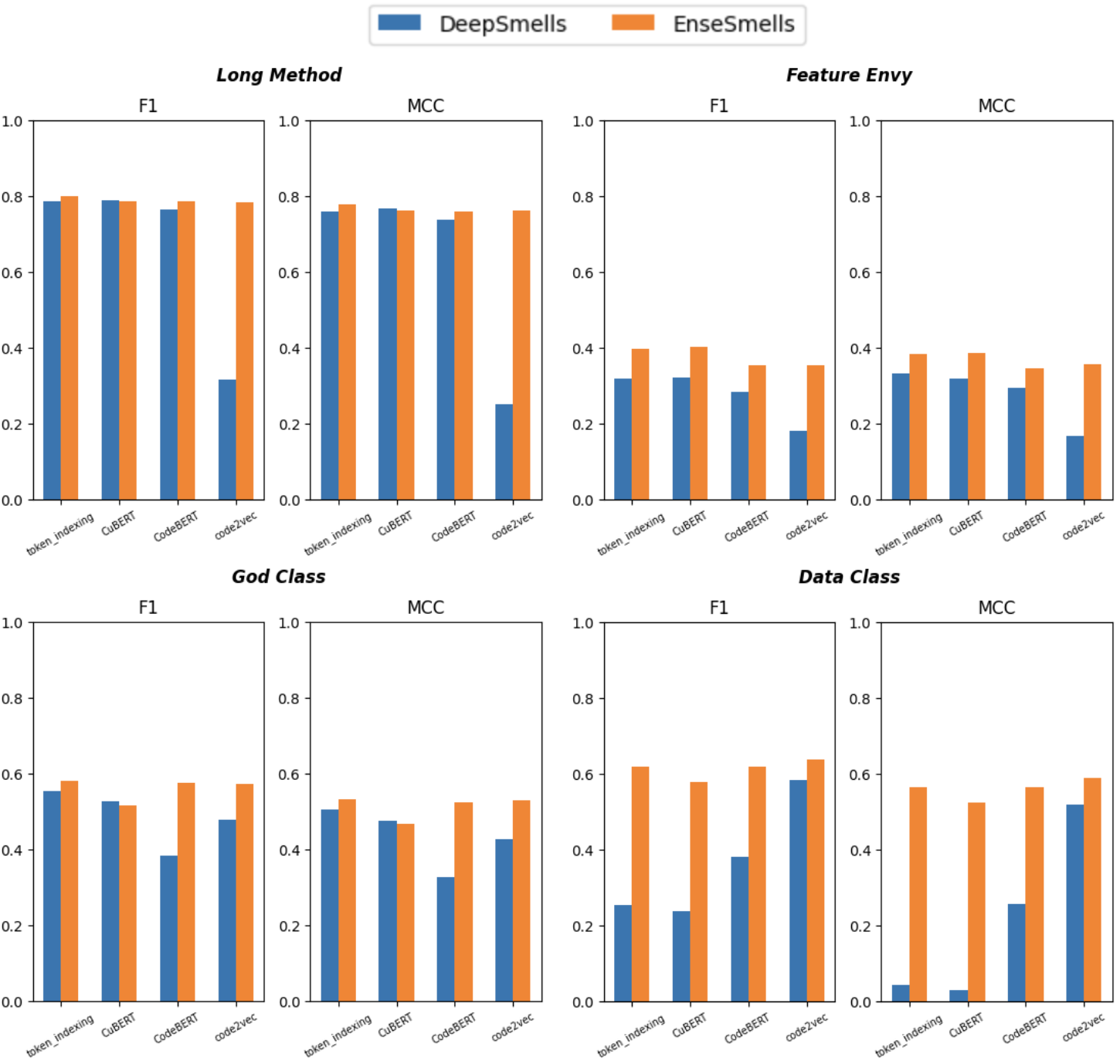}
	\label{fig:role_of_SM}
	\caption{Performance of \ES and \DS across different embedding techniques.} 
	\vspace{-.2cm} 
\end{figure}

Figure~\ref{fig:role_of_SM} compares the performance of \ES and \DS under identical structures and embedding techniques for the semantic module. Notably, \ES incorporates an additional module, the structural model. Analyzing each embedding technique, starting with the \textit{token indexing} method, we observe that \ES outperforms \DS across all four types of smells. Particularly, in the case of DC smell, \ES exhibits significantly higher performance compared to \DS. This trend is similarly observed for \textit{CodeBERT}, with a substantial increase in performance for \ES in GC and DC smells. As for \textit{code2vec}, the same pattern persists, but the considerable outperformance is clearly evident across all smells.

Notably, \textit{CuBERT} presents an exception in LM and GC smell, where \ES slightly trails \DS by less than 1\% in both F1 and MCC metrics. However, in FE smell, \ES outperforms \DS by nearly 10\% in both F1-Score and MCC. The most significant disparity is observed in DC smell, where \ES showcases a substantial performance improvement of approximately 40\% in both F1-Score and MCC compared to \DS.

\vspace{.2cm}
\begin{tcolorbox}[boxrule=0.86pt,left=0.3em, right=0.3em,top=0.1em, bottom=0.05em]
	\textbf{Answer to RQ$_1$.} 
	\rev{Incorporating structural modules enables \ES to obtain a better prediction performance, highlighting the crucial role of software metrics. The performance improvement reaches up to 40\%, depending on the pre-trained model and the type of code smell.}
\end{tcolorbox}
\vspace{-.2cm}

\subsection{\rqsecond}

\begin{figure}[h!]
	\centering
	\vspace{-0.1cm}
	\includegraphics[width=0.80\linewidth]{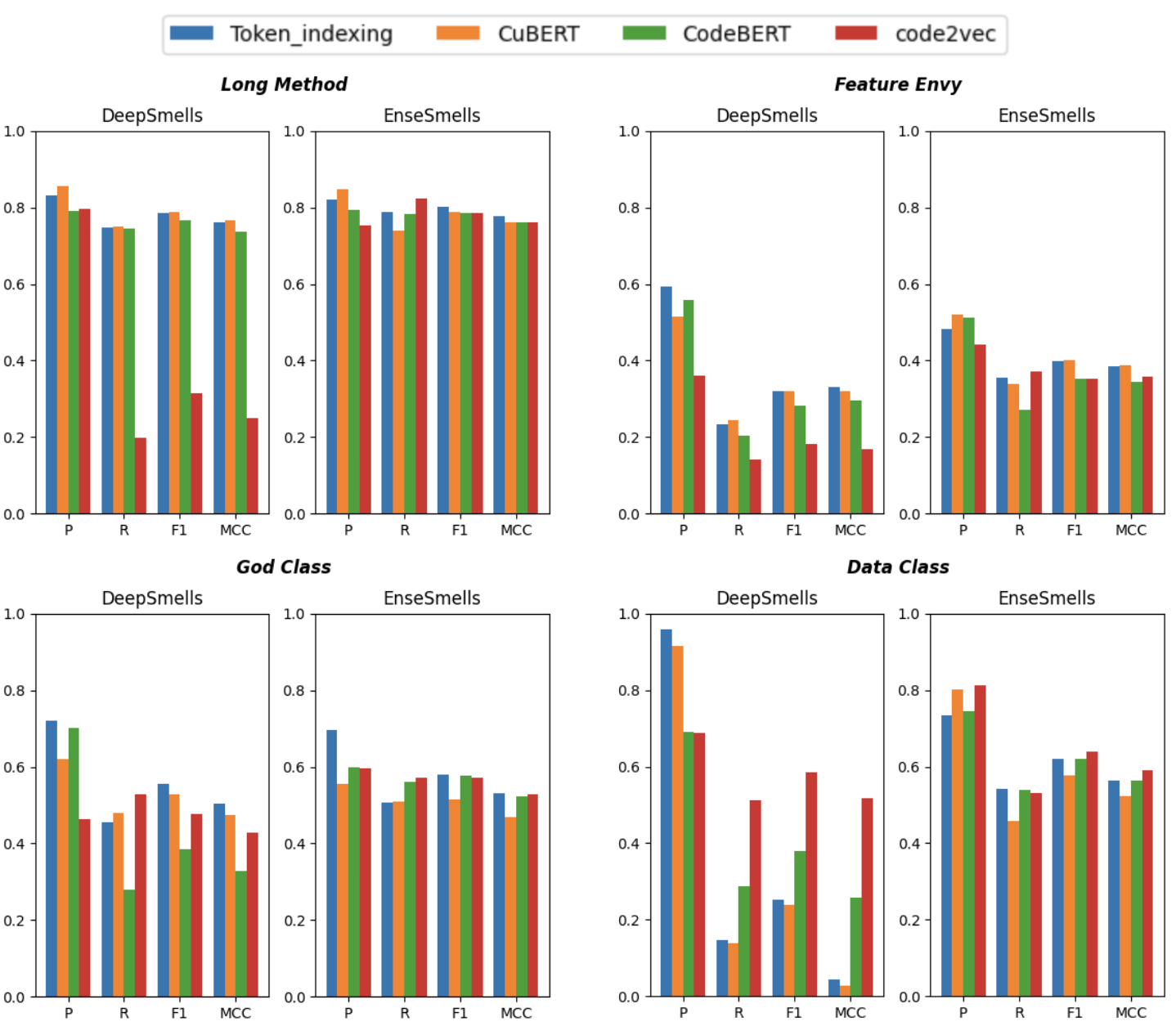}
	\vspace{-0.3cm}
	\caption{Performance of \ES and \DS in various embedding techniques.}
	\label{fig:embedding-smell}
	\vspace{-0.2cm}
\end{figure}

In Figure~\ref{fig:embedding-smell}, we compare the performance of embedding techniques within the same architecture for both \ES and \DS. This study aims to assess the performance of each embedding technique across various smells, examining the performance patterns for each method in relation to each type of smell.

Starting with method-level smells, the use of \textit{code2vec} in both \ES and \DS exhibits the lowest performance compared to other techniques. In contrast, \textit{token indexing} emerges as a highly recommended choice due to its consistently superior performance. Notably, in this context, the consideration of \textit{CuBERT} may be considered, given its potential impact, as it demonstrates only a slightly lower performance than \textit{token indexing}.

Shifting focus to class-level smells, in the case of GC smell, both \ES and \DS employing the \textit{token indexing} embedding technique show the best results compared to other techniques. Transitioning to DC smell, the use of \textit{code2vec} delivers the highest performance, especially in \DS, where it exhibits significant outperformance.


\vspace{.2cm}
\begin{tcolorbox}[boxrule=0.86pt,left=0.3em, right=0.3em,top=0.1em, bottom=0.05em]
	\textbf{Answer to RQ$_2$.} 
	Each pre-trained programming language model possesses a distinct architecture, \anhho{suitable for} addressing specific code smells. Specifically, \textit{token-indexing} is effective for \textit{Long Method} and \textit{God Class}, \textit{CuBERT} is appropriate for \textit{Feature Envy}, and \textit{code2vec} is well-suited for \textit{Data Class}, showcasing the suitability of each model for different types of code smells.
\end{tcolorbox}
\vspace{-.2cm}

\subsection{\rqthird}

Table~\ref{table:ResultsOfMLClassifier} illustrates the performance of \ES in comparison to seven classical machine learning models, which include \textit{Naive Nayes (NB), Nearest Neighbor (NN), Random Forest (RF), Logistic Regression (LR), Classification and Regression Tree (CART), Back Propagation neural networks (BP),} and \textit{Support Vector Machine (SVM)}. These machine learning models exclusively utilize software metrics. The results indicate that \ES outperforms all other models, demonstrating the highest performance across all evaluated models.

\begin{table}[h!]
	\caption{Comparing the performance of seven 
		classifiers on software metrics with \ES model.}
	\label{table:ResultsOfMLClassifier}
	\centering
	\vspace{.3cm}
	\scriptsize
	\begin{tabular}{cccccccccc}
		\Xhline{1pt}
		\multicolumn{1}{c}{\multirow{2}{*}{\textbf{Smell}}} & \multicolumn{1}{c}{\multirow{2}{*}{\textbf{Metric}}} & \multicolumn{7}{c}{\textbf{Machine learning classifier}} & \multicolumn{1}{c}{\multirow{2}{*}{\textbf{\ES}}} \\ \cline{3-9}
		\multicolumn{1}{c}{} & \multicolumn{1}{c}{} & \multicolumn{1}{c}{\textbf{NB}} & \multicolumn{1}{c}{\textbf{NN}} & \multicolumn{1}{c}{\textbf{RF}} & \multicolumn{1}{c}{\textbf{LR}} & \multicolumn{1}{c}{\textbf{CART}} & \multicolumn{1}{c}{\textbf{SVM}} & \multicolumn{1}{c}{\textbf{BP}} & \multicolumn{1}{c}{} \\ \Xhline{1pt}
		
		\multirow{4}{*}{\textbf{LM}} 
		& P & 0.5698 & 0.8048 & 0.8137 & 0.8168 & 0.7595 & 0.7560 & \textbf{0.8477} & 0.8215 \\ 
		& R & \textbf{0.8196} & 0.5310 & 0.6821 & 0.6535 & 0.6429 & 0.6901 & 0.6857 & 0.7873 \\ 
		& F1 & 0.6715 & 0.6360 & 0.7416 & 0.7248 & 0.6943 & 0.7199 & 0.7566 & \textbf{0.8016} \\ 
		& MCC & 0.6339 & 0.6173 & 0.7150 & 0.6994 & 0.6625 & 0.6874 & 0.7347 & \textbf{0.7780} \\ \Xhline{1pt}
		
		\multirow{4}{*}{\textbf{FE}} 
		& P & 0.1162 & 0.1265 & 0.3000 & 0.2000 & 0.4611 & 0.2246 & 0.2126 & \textbf{0.5215} \\ 
		& R & \textbf{0.7440} & 0.1220 & 0.0462 & 0.0736 & 0.2099 & 0.1220 & 0.2451 & 0.3393 \\ 
		& F1 & 0.1993 & 0.1226 & 0.0800 & 0.1076 & 0.2784 & 0.1545 & 0.2181 & \textbf{0.4025} \\ 
		& MCC & 0.2453 & 0.0987 & 0.1084 & 0.1068 & 0.2911 & 0.1441 & 0.1978 & \textbf{0.3868} \\  \Xhline{1pt}
		
		\multirow{4}{*}{\textbf{GC}} 
		& P & 0.1620 & 0.5415 & \textbf{0.7442} & 0.6231 & 0.5296 & 0.6109 & 0.5836 & 0.6220 \\ 
		& R & \textbf{0.8308} & 0.3943 & 0.3662 & 0.2744 & 0.4505 & 0.4011 & 0.5494 & 0.5867 \\ 
		& F1 & 0.2709 & 0.4552 & 0.4895 & 0.3760 & 0.4845 & 0.4809 & 0.5656 & \textbf{0.5851} \\ 
		& MCC & 0.1280 & 0.3944 & 0.4761 & 0.3570 & 0.4166 & 0.4339 & 0.5028 & \textbf{0.5429} \\  \Xhline{1pt}
		
		\multirow{4}{*}{\textbf{DC}} 
		& P & 0.1644 & 0.5415 & 0.6849 & 0.5980 & 0.5525 & 0.6343 & 0.5839 & \textbf{0.8123} \\ 
		& R & \textbf{0.8205} & 0.3943 & 0.3979 & 0.2816 & 0.4263 & 0.3838 & 0.5393 & 0.5314 \\ 
		& F1 & 0.2734 & 0.4552 & 0.5019 & 0.3803 & 0.4808 & 0.4772 & 0.5573 & \textbf{0.6393} \\ 
		& MCC & 0.1291 & 0.3944 & 0.4703 & 0.3541 & 0.4184 & 0.4366 & 0.4971 & \textbf{0.5907} \\  \Xhline{1pt}
	\end{tabular}
\end{table}

Specifically, for the LM code smell, the F1-Score and MCC of \ES are 80.16\% and 77.80\% higher than those of BP, with improvements of 4.50\% and 4.33\%, respectively. Notably, the precision value of BP is 2.62\% higher than \ES but comes at the cost of a significant reduction in recall by 10.16\%. Additionally, while the recall value of NB achieves the highest at 81.96\%, it is higher than \ES; however, the precision, F1-Score, and MCC values are significantly lower by 25.17\%, 13.01\%, and 14.41\%, respectively. 

Regarding the FE code smell, when comparing \ES to CART, which is considered the best machine learning classifier for this type of code smell, \ES outperforms in all four evaluation metrics, showing improvements of 6.04\%, 12.94\%, 12.41\%, and 9.57\% in precision, recall, F1-Score, and MCC, respectively. The scenario is significantly different when comparing NB and \ES. While NB achieves the highest precision value among these models at 74.40\%, its recall value is only 11.62\%, highlighting its weakness in handling the high imbalanced dataset. 

Turning to GC code smell, while BP exhibits the best overall performance among these ML classifiers, when compared to \ES, our model demonstrates outperformance in all four metrics. Furthermore, NB and RF are two models with the highest values in recall and precision, but they also exhibit significant reductions in precision and recall, respectively. Consequently, this leads to a significant decrease in both F1-Score and MCC, ranging from 6.68\% to 41.49\%.

As for the remaining code smell, DC, when comparing \ES to BP, which demonstrates good performance across almost all types of code smells including this one, our model reveals significant differences in precision, F1-Score, and MCC, corresponding to 22.84\%, 8.2\%, and 9.36\%. Additionally, when comparing \ES to NB, which boasts the highest recall value at 82.05\%, the precision value of only 16.44\% highlights its weakness in handling highly imbalanced datasets leading to a remarkable decrease in both F1-Score and MCC, with a deviation of 36.59\% and 46.16\% lower than \ES, respectively.



\vspace{.2cm}
\begin{tcolorbox}[boxrule=0.86pt,left=0.3em, right=0.3em,top=0.1em, bottom=0.05em]
	\textbf{Answer to RQ$_3$.} \anhho{\ES outperforms all traditional ML classifiers that build prediction models using \textit{structural} features. In terms of F1-Score, \ES achieves superior performance ranging from 1.95\% to 12.41\% compared to the best ML classifier for each code smell. For MCC, the improvement ranges from 4.01\% to 14.15\%.}
\end{tcolorbox}
\vspace{-.2cm}

\subsection{\rqfourth}

Tables~\ref{table:ResultsOfEnsemble} and~\ref{table:ResultsOfEASE} present the optimal performance of the \ES and \DS models, respectively, evaluated using different code embedding techniques, including \textit{token-indexing, CuBERT, CodeBERT}, and \textit{code2vec}, each under their respective optimal network configurations. 

\begin{table}[h!]	
	\caption{Performance of different embedding techniques in \ES architecture under optimal configurations.}
	\label{table:ResultsOfEnsemble}
	\centering
	\scriptsize
	\vspace{0.3cm}
	\begin{tabular}{clcccc}
		\Xhline{1pt}
		\multicolumn{1}{c}{\multirow{2}{*}{\textbf{Smell}}} & \multicolumn{1}{c}{\multirow{2}{*}{\textbf{Model}}} & \multicolumn{4}{c}{\textbf{Evaluation metric}} \\ 
		\cline{3-6}
		\multicolumn{1}{c}{} & \multicolumn{1}{c}{} & \multicolumn{1}{c}{\textbf{P}} & \multicolumn{1}{c}{\textbf{R}} & \multicolumn{1}{c}{\textbf{F1}} & \multicolumn{1}{c}{\textbf{MCC}} \\ 
		\Xhline{1pt}
		\multirow{4}{*}{\textbf{LM}} 
		& \textbf{$\ES_{token-indexing}$}  & 0.8215 & 0.7873 & \textbf{0.8016} & \textbf{0.7780} \\ 
		& \textbf{$\ES_{CuBERT}$} & \textbf{0.8482} & 0.7391 & 0.7877 & 0.7616 \\ 
		& \textbf{$\ES_{CodeBERT}$} & 0.7938 & 0.7824 & 0.7857 & 0.7611 \\
		& \textbf{$\ES_{code2vec}$} & 0.7526 & \textbf{0.8242} & 0.7841 & 0.7618 \\ 
		\Xhline{1pt}
		\multirow{4}{*}{\textbf{FE}}  
		& \textbf{$\ES_{token-indexing}$} & 0.4821 & 0.3554 & 0.3982 & 0.3849 \\ 
		& \textbf{$\ES_{CuBERT}$} & \textbf{0.5215} & \textbf{0.3393} & \textbf{0.4025} & \textbf{0.3868} \\ 
		& \textbf{$\ES_{CodeBERT}$} & 0.5128 & 0.2718 & 0.3527 & 0.3449 \\
		& \textbf{$\ES_{code2vec}$} & 0.4423 & 0.3727 & 0.3538 & 0.3568 \\ 
		\Xhline{1pt}
		\multirow{4}{*}{\textbf{GC}}  
		& \textbf{$\ES_{token-indexing}$} & \textbf{0.6962} & 0.5077 & \textbf{0.5801} & \textbf{0.5318} \\ 
		& \textbf{$\ES_{CuBERT}$} & 0.5565 & 0.5105 & 0.5153 & 0.4675 \\ 
		& \textbf{$\ES_{CodeBERT}$} & 0.5993 & 0.5607 & 0.5764 & 0.5241 \\
		& \textbf{$\ES_{code2vec}$} & 0.5947 & \textbf{0.5724} & 0.5729 & 0.5288 \\ 
		\Xhline{1pt}
		\multirow{4}{*}{\textbf{DC}}  
		& \textbf{$\ES_{token-indexing}$} & 0.7340 & \textbf{0.5427} & 0.6203 & 0.5638 \\ 
		& \textbf{$\ES_{CuBERT}$} & 0.8027 & 0.4567 & 0.5772 & 0.5238 \\ 
		& \textbf{$\ES_{CodeBERT}$} & 0.7447 & 0.5390 & 0.6199 & 0.5648 \\
		& \textbf{$\ES_{code2vec}$} & \textbf{0.8123} & 0.5314 & \textbf{0.6393} & \textbf{0.5907} \\ \Xhline{1pt}
	\end{tabular}
\end{table}

\begin{table}[h!]
	\scriptsize
	\caption{\anhho{Performance of different embedding techniques in \DS architecture under optimal configurations.}}
	\label{table:ResultsOfEASE}
	\centering
	\vspace{.3cm}
	\begin{tabular}{clcccc}
		\Xhline{1pt}
		\multicolumn{1}{c}{\multirow{2}{*}{\textbf{Smell}}} & \multicolumn{1}{c}{\multirow{2}{*}{\textbf{Model}}} & \multicolumn{4}{c}{\textbf{Metric}} \\ 
		\cline{3-6}
		\multicolumn{1}{c}{} & \multicolumn{1}{c}{} & \multicolumn{1}{c}{\textbf{P}} & \multicolumn{1}{c}{\textbf{R}} & \multicolumn{1}{c}{\textbf{F1}} & \multicolumn{1}{c}{\textbf{MCC}} \\ 
		\Xhline{1pt}
		\multirow{5}{*}{\textbf{LM}} 
		& \textbf{$\DS_{token-indexing}$}~\citep{ho2023fusion} & 0.8327 & 0.7470 & 0.7865 & 0.7607 \\ 
		& \textbf{$\DS_{CuBERT}$} & \textbf{0.8560} & \textbf{0.7490} & \textbf{0.7888} & \textbf{0.7673} \\ 
		& \textbf{$\DS_{CodeBERT}$} & 0.7903 & 0.7448 & 0.7661 & 0.7373  \\
		& \textbf{$\DS_{code2vec}$} & 0.7974 & 0.1982 & 0.3152 & 0.2509 \\ 
		\Xhline{1pt}
		
		\multirow{5}{*}{\textbf{FE}}  
		& \textbf{$\DS_{token-indexing}$}~\citep{ho2023fusion} & \textbf{0.5936} & 0.2322 & 0.3198 & \textbf{0.3318} \\ 
		& \textbf{$\DS_{CuBERT}$} & 0.5143 & \textbf{0.2439} & \textbf{0.3203} & 0.3201 \\ 
		& \textbf{$\DS_{CodeBERT}$} & 0.5590 & 0.2048 & 0.2833 & 0.2947  \\
		& \textbf{$\DS_{code2vec}$} & 0.3603 & 0.1422 & 0.1815 & 0.1683 \\ 
		\Xhline{1pt}
		
		\multirow{5}{*}{\textbf{GC}}  
		& \textbf{$\DS_{token-indexing}$}~\citep{ho2023fusion} & \textbf{0.7193} & 0.4559 & \textbf{0.5542} & \textbf{0.5049} \\ 
		& \textbf{$\DS_{CuBERT}$} & 0.6191 & \textbf{0.4809} & 0.5277 & 0.4747 \\ 
		& \textbf{$\DS_{CodeBERT}$} & 0.7007 & 0.2798 & 0.3845 & 0.3270  \\
		& \textbf{$\DS_{code2vec}$} & 0.4647 & 0.5293 & 0.4780 & 0.4272 \\
		\Xhline{1pt}
		
		\multirow{5}{*}{\textbf{DC}}  
		& \textbf{$\DS_{token-indexing}$}~\citep{ho2023fusion} & \textbf{0.9579} & 0.1476 & 0.2531 & 0.0439 \\ 
		& \textbf{$\DS_{CuBERT}$} & 0.9158 & 0.1389 & 0.2382 & 0.0281 \\ 
		& \textbf{$\DS_{CodeBERT}$} & 0.6916 & 0.2868 & 0.3802 & 0.2573  \\
		& \textbf{$\DS_{code2vec}$} & 0.6883 & \textbf{0.5132} & \textbf{0.5839} & \textbf{0.5185} \\ 
		\Xhline{1pt}
	\end{tabular}
\end{table}

\anhho{Table~\ref{table:ESandDS} demonstrates the performance of \DS and \ES under their best configurations, showing that \ES outperforms \DS. For the LM code smell, \ES achieves improvements of 1.28\% in F1-Score and 1.07\% in MCC. Although \DS exhibits a 3.45\% higher precision, this comes at the expense of a 3.83\% lower recall. Similarly, for the GC code smell, \ES surpasses \DS by 3.09\% in F1-Score and 3.8\% in MCC. While \DS achieves a higher precision by 9.73\%, its recall is 13.09\% lower compared to \ES.}

\anhho{For the remaining code smells, \ES shows superior performance across all evaluation metrics. Notably, \ES outperforms \DS by 8.27\% in F1-Score and 5.5\% in MCC for the FE code smell, and by 7.22\% in F1-Score and 5.5\% in MCC for the DC code smell.}

\begin{table}[h!]
	\scriptsize
	\caption{\anhho{Comparison of \DS and \ES models with optimal configurations.}}
	\label{table:ESandDS}
	\centering
	\vspace{.3cm}
	\begin{tabular}{clcccc}
		\Xhline{1pt}
		\multicolumn{1}{c}{\multirow{2}{*}{\textbf{Smell}}} & \multicolumn{1}{c}{\multirow{2}{*}{\textbf{Model}}} & \multicolumn{4}{c}{\textbf{Metric}} \\ 
		\cline{3-6}
		\multicolumn{1}{c}{} & \multicolumn{1}{c}{} & \multicolumn{1}{c}{\textbf{P}} & \multicolumn{1}{c}{\textbf{R}} & \multicolumn{1}{c}{\textbf{F1}} & \multicolumn{1}{c}{\textbf{MCC}} \\ 
		\Xhline{1pt}
		\multirow{2}{*}{\textbf{LM}}
		& \textbf{\DS} & \textbf{0.8560} & 0.7490 & 0.7888 & 0.7673 \\ 
		& \textbf{\ES} & 0.8215 & \textbf{0.7873} & \textbf{0.8016} & \textbf{0.7780} \\ 
		\Xhline{1pt}
		
		\multirow{2}{*}{\textbf{FE}}  
		& \textbf{\DS} & 0.5936 & 0.2322 & 0.3198 & 0.3318 \\ 
		& \textbf{\ES} & \textbf{0.5215} & \textbf{0.3393} & \textbf{0.4025} & \textbf{0.3868} \\ 
		\Xhline{1pt}
		
		\multirow{2}{*}{\textbf{GC}}  
		& \textbf{\DS} & \textbf{0.7193} & 0.4559 & 0.5542 & 0.5049 \\ 
		& \textbf{\ES} & 0.6220 & \textbf{0.5867} & \textbf{0.5851} & \textbf{0.5429} \\ 
		\Xhline{1pt}
		
		\multirow{2}{*}{\textbf{DC}}  
		& \textbf{\DS} & 0.6883 & 0.5132 & 0.5839 & 0.5185 \\ 
		& \textbf{\ES} & \textbf{0.8123} & \textbf{0.5314} & \textbf{0.6393} & \textbf{0.5907} \\ 
		\Xhline{1pt}
	\end{tabular}
\end{table}

\begin{table}[h!]
	\scriptsize
	\caption{Comparison with baseline models.}
	\label{table:ResultOfSOTA}
	\centering
	\vspace{.3cm}
	\begin{tabular}{clcccc}
		\Xhline{1pt}
		\multicolumn{1}{c}{\multirow{2}{*}{\textbf{Smell}}} & \multicolumn{1}{c}{\multirow{2}{*}{\textbf{Model}}} & \multicolumn{4}{c}{\textbf{Metric}} \\ 
		\cline{3-6}
		\multicolumn{1}{c}{} & \multicolumn{1}{c}{} & \multicolumn{1}{c}{\textbf{P}} & \multicolumn{1}{c}{\textbf{R}} & \multicolumn{1}{c}{\textbf{F1}} & \multicolumn{1}{c}{\textbf{MCC}} \\ 
		\Xhline{1pt}
		\multirow{5}{*}{\textbf{LM}} 
		& \textbf{ML\_CuBERT}~\citep{kovavcevic2022automatic} & 0.6933 & 0.8142 & 0.7481 & 0.7182 \\ 
		& \textbf{AE-Dense}~\citep{sharma2021code} & 0.7432 & 0.8274 & 0.7804 & 0.7548 \\ 
		& \textbf{AE-CNN}~\citep{sharma2021code} & 0.7363 & 0.8315 & 0.7728 & 0.7500 \\
		& \textbf{AE-LSTM}~\citep{sharma2021code} & 0.6680 & \textbf{0.8480} & 0.7471 & 0.7187 \\ 
		& \textbf{\ES} & \textbf{0.8215} & 0.7873 & \textbf{0.8016} & \textbf{0.7780} \\ 
		\Xhline{1pt}
		
		\multirow{5}{*}{\textbf{FE}}  
		& \textbf{ML\_CuBERT}~\citep{kovavcevic2022automatic} & 0.1223 & \textbf{0.8033} & 0.2121 & 0.2683 \\ 
		& \textbf{AE-Dense}~\citep{sharma2021code} & 0.2325 & 0.6295 & 0.3356 & 0.3509 \\ 
		& \textbf{AE-CNN}~\citep{sharma2021code} & 0.2704 & 0.5333 & 0.3500 & 0.3478 \\
		& \textbf{AE-LSTM}~\citep{sharma2021code} & 0.1364 & 0.7987 & 0.2329 & 0.2888 \\ 
		& \textbf{\ES} & \textbf{0.5215} & 0.3393 & \textbf{0.4025} & \textbf{0.3868} \\ 
		\Xhline{1pt}
		
		\multirow{5}{*}{\textbf{GC}}  
		& \textbf{ML\_CuBERT}~\citep{kovavcevic2022automatic} & 0.4580 & 0.5763 & 0.5092 & 0.4492 \\ 
		& \textbf{AE-Dense}~\citep{sharma2021code} & 0.5495 & 0.6444 & 0.5751 & 0.5306 \\ 
		& \textbf{AE-CNN}~\citep{sharma2021code} & 0.5334 & \textbf{0.6622} & 0.5775 & 0.5311 \\
		& \textbf{AE-LSTM}~\citep{sharma2021code} & 0.5227 & 0.6490 & 0.5706 & 0.5200 \\ 
		& \textbf{\ES} & \textbf{0.6220} & 0.5867 & \textbf{0.5851} & \textbf{0.5429} \\ 
		\Xhline{1pt}
		
		\multirow{5}{*}{\textbf{DC}}  
		& \textbf{ML\_CuBERT}~\citep{kovavcevic2022automatic} & 0.4953 & 0.2853 & 0.3615 & 0.3081 \\ 
		& \textbf{AE-Dense}~\citep{sharma2021code} & 0.1559 & 0.8576 & 0.2613 & 0.1068 \\ 
		& \textbf{AE-CNN}~\citep{sharma2021code} & 0.1489 & \textbf{0.9184} & 0.2561 & 0.0977 \\
		& \textbf{AE-LSTM}~\citep{sharma2021code} & 0.1532 & 0.8328 & 0.2567 & 0.0922 \\ 
		& \textbf{\ES} & \textbf{0.8123} & 0.5314 & \textbf{0.6393} & \textbf{0.5907} \\ 
		\Xhline{1pt}
	\end{tabular}
\end{table}

Table~\ref{table:ResultOfSOTA} presents a comparison between \ES and the state-of-the-art approach, \textit{ML\_CuBERT} \citep{kovavcevic2022automatic}, and existing approaches that include various auto-encoder variants~\citep{sharma2021code}. The table demonstrates that our proposed model outperforms the other models by all evaluation metrics. Especially for LM, \ES exhibits the lowest recall value at 78.73\%, which is lower than the others by 2.69\% to 6.07\%. However, it boasts the highest precision value, surpassing the others by 7.83\% to 15.35\%. As a result, the overall performance demonstrates the highest F1-Score and MCC, with significant differences compared to the other models. Regarding the FE code smell, following a similar pattern LM, \ES showcases an overall outperformance, with substantial differences compared to ML\_CuBERT and the variants of autoencoder. The F1-Score varies from 5.25\% to 19.04\%, while the MCC differs from 3.59\% to 11.85\%.

In relation to the GC code smell, compared to AE-CNN, which shows the best performance among the baseline models with values of 53.34\% (precision), 66.22\% (recall), 57.75\% (F1-Score), and 53.133\% (MCC), \ES exhibits a slightly higher in F1-Score and MCC by 0.76\% and 1.18\%, respectively. Notably, the precision value of \ES is 8.86\% higher than AE-CNN, while the recall value is only 7.55\% lower.

\anhho{The remaining code smell is the DC}. The autoencoder variants exhibit a weakness in dealing with highly imbalanced datasets, showcasing very high recall values of up to 91.84\%, but the precision values hover around 15\%, resulting in the lowest overall performance. Moreover, \ES outperforms ML\_CuBERT in all four evaluation metrics by 31.7\% (precision), 24.61\% (recall), 27.78\% (F1-Score), and 28.26\% (MCC).

\vspace{.2cm}
\begin{tcolorbox}[boxrule=0.86pt,left=0.3em, right=0.3em,top=0.1em, bottom=0.05em]
	\textbf{Answer to RQ$_4$.} \rebuttal{
		\ES demonstrates superior prediction performance for all four types of code smells compared to the state-of-the-art baselines. It gains a better accuracy compared to that of \DS, with improvements ranging from approximately 5\% to 10\%. 
		The MCC (overall performance) surpasses ML\_CuBERT by approximately 5.98\% \rev{to} 28.26\% and the best Autoencoder variants by 1.18\% \rev{to} 48.39\% across the code smell categories.}
\end{tcolorbox}
\vspace{-.2cm}


\section{Threat to validity}\label{sec:threat}

\rebuttal{We anticipate that there are the following threats to the validity of our findings.}

\begin{itemize}[leftmargin=*]
\vspace{-0.3cm}\item \textbf{Internal validity.} \anhho{This concerns the extent to which our evaluation reflects real-world scenarios. In our work,  we used existing datasets~\citep{madeyski2020mlcq} curated and classified by human experts. Additionally, we assigned labels to each code snippet based on majority-based approach. However, these labels may not be universally accepted, which could impact the overall dataset quality and, consequently, the accuracy of our predictions. Furthermore, the dataset also exhibits class imbalance, mirroring real-world distributions but potentially biasing the model towards more common classes. To address this, we applied class-weight adjustments during training and used metrics designed for imbalanced data, ensuring a fair evaluation across all classes.}
\vspace{-0.3cm}\item \textbf{External validity.} This concerns the generalizability of the findings beyond the scope of this study. We attempted to mitigate threats by evaluating with different experimental configurations to simulate real-world scenarios. The findings of our work might apply only to the considered datasets. For other datasets, additional empirical evidence is required before reaching a final conclusion.
\vspace{-0.3cm}\item \textbf{Construct validity.} This dimension relates to how we set up our experiments to compare \ES with the baseline models. To ensure a fair comparison, we used the original implementations provided by the authors of the baseline models, maintaining their internal structures. 
Furthermore, our method involved measuring various software metrics using well-accepted open-source tools in our field. It is crucial to note that our study focused on the extensive MLCQ dataset, comprising a substantial number of projects (792). In such a vast dataset, there is a possibility of errors in metric calculations due to parsing issues. During our error analysis, our domain expert identified a few instances where unique word counts were miscalculated. 
Nevertheless, we have confidence that the use of established tools widely recognized in the research community helps minimize the impact of these potential errors.
\end{itemize}



\section{Conclusion}\label{sec:Conclusion}
This paper introduces \ES, an innovative approach for detecting four types of code smells - two at the method level (\textit{Long Method} and \textit{Feature Envy}) and two at the class level (\textit{God Class} and \textit{Data Class}). Our methodology leverages a unique ensemble of two feature types: \textit{semantic features} derived from novel pre-trained programming language models and \textit{structural features} obtained through object-oriented metrics extracted by a static analysis tool. Building upon the foundation of our sophisticated model in a previous work, \DS, we conduct extensive experiments to address five research questions and assess the effectiveness of \ES. The results demonstrate that \ES outperforms existing methods, achieving state-of-the-art performance on the MLCQ dataset \anhho{with improvements ranging from 5.98\% to 28.26\%, depending on the type of code smell.}
Our future work involves expanding the application of this approach to additional code smells and identifying areas for improvement to further enhance its capabilities.

\section*{Acknowledgment}
\rebuttal{Our research has been funded by Hanoi University of Science and Technology (HUST), Vietnam under project number T2023-PC-002. 
This work has been supported by the COmmunity-Based Organized Littering (COBOL) national research project funded by the MUR under the PRIN 2022 PNRR program (nr. P20224K9EK), by the ``Progetto PE 0000020 CHANGES,  PNRR Missione 4 Componente 2 Investimento 1.3'' funded by EU - NextGenerationEU, by the European HORIZON-KDT-JU research project MATISSE ``Model-based engineering of Digital Twins for early verification and validation of Industrial Systems'', HORIZON-KDT-JU-2023-2-RIA, Proposal number:  101140216-2, KDT232RIA\_00017, and by the European Union - NextGenerationEU under the Italian Ministry of University and Research (MUR) National Innovation Ecosystem grant ECS00000041 - VITALITY – CUP: D13C21000430001. We acknowledge the Italian ``PRIN 2022'' project TRex-SE: \emph{``Trustworthy Recommenders for Software Engineers,''} grant n. 2022LKJWHC. 
We thank the anonymous reviewers for their useful comments and suggestions that helped us improve our manuscript.}

\balance

\newpage

\centering \appendix \section{The Information of Software Metrics at Method and Class Level}
\label{sec:Appendix}

\begin{table}[ht]
	\caption{The abbreviation and software metric description for method level.}
	\label{table:DescriptionMethodLevel}
	\tiny
	\centering
	\begin{tabularx}{\textwidth}{>{\hsize=.25\hsize}X>{\hsize=.75\hsize}X}
		\Xhline{0.3pt}
		\textbf{Abbreviation}          & \textbf{Software Metric Description}          \\
		\endhead
		\Xhline{0.3pt}
		\hline
		LOC                            & Lines of code (excluding empty lines and comments). \\
		\hline
		CBO                            & Coupling between objects: measures the number of classes or methods a given method depends on, considering parameters, return types, local variables, and method calls, excluding dependencies on Java standard libraries (e.g., java.lang.String). \\
		\hline
		WMC                            & Weighted method complexity: counts the number of branch instructions (e.g., decision points like if, for, while) within methods, based on McCabe's complexity. \\
		\hline
		RFC                            & Response for a class: counts the unique method invocations within a method. \\
		\hline
		modifiers                      & The modifiers of methods, including public, private, abstract, protected, and native. \\
		\hline
		constructor                    & Indicates whether the method is a constructor. \\
		\hline
		logStatementsQty               & The number of log statements. \\
		\hline
		returnsQty                     & The number of return instructions. \\
		\hline
		variablesQty                   & The number of variables declared. \\
		\hline
		parametersQty                  & The number of parameters. \\
		\hline
		methodsInvokedQty              & The total number of methods directly invoked, including local and indirect local invocations. \\
		\hline
		methodsInvokedLocalQty         & The number of methods invoked locally within the method. \\
		\hline
		methodsInvokedIndirectLocalQty & The number of indirectly invoked methods within the method. \\
		\hline
		loopQty                        & The number of loops (e.g., for, while, do-while) within methods. \\
		\hline
		comparisonsQty                 & The number of comparison operations (e.g., ==, !=, <, >) within methods. \\
		\hline
		tryCatchQty                    & The number of try/catch blocks within methods. \\
		\hline
		parenthesizedExpsQty           & The number of expressions inside parentheses within methods. \\
		\hline
		stringLiteralsQty              & The number of string literals used in methods. \\
		\hline
		numbersQty                     & The number of numerical literals (e.g., int, long, double, float) used in methods. \\
		\hline
		assignmentsQty                 & The number of assignment operations within methods. \\
		\hline
		mathOperationsQty              & The number of mathematical operations (e.g., +, -, *, /) within methods. \\
		\hline
		maxNestedBlocksQty             & The maximum depth of nested blocks (e.g., loops, conditionals) within methods. \\
		\hline
		anonymousClassesQty            & The number of anonymous classes used within methods. \\
		\hline
		innerClassesQty                & The number of inner classes defined within methods. \\
		\hline
		lambdasQty                     & The number of lambda expressions used in methods. \\
		\hline
		uniqueWordsQty                 & The number of unique words (e.g., identifiers, keywords) used in methods. \\
		\hline
		\Xhline{0.3pt}
	\end{tabularx}
\end{table}

\begin{table}[]
	\caption{The abbreviation and software metric description for class level.}
	\label{table:DescriptionClassLevel}
	\tiny
	\centering
	\begin{tabularx}{\textwidth}{>{\hsize=.2\hsize}X>{\hsize=.8\hsize}X}
		\Xhline{0.3pt}
		\textbf{Abbreviation}           & \textbf{Software Metric Description}              \\
		\endhead
		\Xhline{0.3pt}
		\hline
		CBO                    & Coupling between objects measures the number of dependencies a class has, considering all types used within the class (fields, method return types, variables, etc.), excluding dependencies on Java standard libraries (e.g., java.lang.String). \\
		\hline
		DIT                    & Depth inheritance tree counts the number of parent classes a class has. The minimum DIT is 1 (all classes inherit \texttt{java.lang.Object}). If a class depends on an external dependency (e.g., a JAR file), the depth is counted as 2.\\
		\hline
		WMC                    & Weight method class measures the complexity of a class by counting the number of branch instructions (e.g., decision points like if, for, while) within its methods, based on McCabe's complexity.\\
		\hline
		TCC                    & Tight class cohesion measures the cohesion of a class (within a range of 0 to 1) based on direct connections between visible methods, where methods or their invocation trees access the same class variable. \\
		\hline
		LCC                    & Loose class cohesion extends TCC by including indirect connections between visible classes, ensuring LCC $\geq$ TCC. \\
		\hline
		LCOM                   & Lack of cohesion of methods measures the cohesion within a class. Higher cohesion indicates a well-structured class, while lower cohesion suggests the class may handle multiple responsibilities. \\
		\hline
		LOC                    & Lines of code (excluding empty lines and comments). \\
		\hline
		NOSI                   & The number of static invocations counts the number of invocations to static methods. \\
		\hline
		RFC                    & Response for a class counts unique method invocations in a class. \\
		\hline
		abstractMethodsQty     & The number of abstract methods. \\
		\hline
		anonymousClassesQty    & The number of anonymous classes. \\
		\hline
		assignmentsQty         & The number of assignments. \\
		\hline
		comparisonsQty         & The number of comparisons. \\
		\hline
		defaultFieldsQty       & The number of default fields. \\
		\hline
		defaultMethodsQty      & The number of default methods. \\
		\hline
		finalFieldsQty         & The number of final fields. \\
		\hline
		finalMethodsQty        & The number of final methods. \\
		\hline
		innerClassesQty        & The number of inner classes. \\
		\hline
		lambdasQty             & The number of lambda expressions. \\
		\hline
		logStatementsQty       & The number of log statements. \\
		\hline
		loopQty                & The number of loops (i.e., for, while, do while). \\
		\hline
		mathOperationsQty      & The number of math operations. \\
		\hline
		maxNestedBlocksQty     & The number of maximum nested blocks. \\
		\hline
		modifiers              & The modifiers of classes include public, abstract, private, protected, or native. \\
		\hline
		privateFieldsQty       & The number of private fields. \\
		\hline
		privateMethodsQty      & The number of private methods. \\
		\hline
		protectedFieldsQty     & The number of protected fields. \\
		\hline
		protectedMethodsQty    & The number of protected methods. \\
		\hline
		publicFieldsQty        & The number of public fields. \\
		\hline
		publicMethodsQty       & The number of public methods. \\
		\hline
		returnQty              & The number of return instructions. \\
		\hline
		staticFieldsQty        & The number of static fields. \\
		\hline
		staticMethodsQty       & The number of static methods. \\
		\hline
		stringLiteralsQty      & The number of string literals. \\
		\hline
		synchronizedFieldsQty  & The number of synchronized fields. \\
		\hline
		synchronizedMethodsQty & The number of synchronized methods. \\
		\hline
		totalFieldsQty         & The total number of fields. \\
		\hline
		totalMethodsQty        & The total number of methods. \\
		\hline
		tryCatchQty            & The number of try/catch blocks. \\
		\hline
		visibleFieldsQty       & The number of visible fields. \\
		\hline
		numbersQty             & The number of numerical literals (e.g., int, long, double, float). \\
		\hline
		parenthesizedExpsQty   & The number of expressions inside parenthesis. \\
		\hline
		uniqueWordsQty         & The number of unique words. \\
		\hline
		variablesQty           & The number of variables. \\
		
		\Xhline{0.3pt}
	\end{tabularx}
\end{table}

\end{document}